\documentclass[twocolumn,prb,aps]{revtex4}

\usepackage{graphicx}
\usepackage{dcolumn}
\usepackage{bm}

\begin{document}

\preprint{submitted to Phys. Rev. B}

\title{Observation of room-temperature spontaneous phase segregation in overdoped 
{\rm Bi$_2$Sr$_2$CaCu$_2$O$_{8+x}$} 
single crystals}

\author{X. S. Wu}
\author{L. Lu}%
\altaffiliation[Correspondence author, ]{email: lilu@aphy.iphy.ac.cn.}
\author{D. L. Zhang}
\affiliation{%
Key Laboratory of Extreme Conditions Physics, 
Institute of Physics \& Center for Condensed Matter Physics\\ 
Chinese Academy of Sciences, Beijing 100080, P. R. China
}%

\author{Y. Xuan}
\author{H. J. Tao}
\affiliation{
National Laboratory for Superconductivity, 
Institute of Physics \& Center for Condensed Matter Physics\\ 
Chinese Academy of Sciences, Beijing 100080, P. R. China
}%


\begin{abstract}
The occurrence and development of phase 
inhomogeneity in {\rm Bi$_2$Sr$_2$CaCu$_2$O$_{8+x}$} 
single crystals have been investigated 
by measuring their superconducting transition 
patterns using a high-sensitivity ac magnetometer. We find that 
the overdoped {\rm Bi$_2$Sr$_2$CaCu$_2$O$_{8+x}$} single crystals, 
even if they were single-phased immediately after a high-temperature 
annealing process, show progressively pronounced multiple superconducting 
transitions with time as they were kept at room temperature in a dry air 
atmosphere. The results reveal that the oxygen dopants in the crystal
are still mobile at room temperature. These dopants tend to re-arrange and 
thus form different phases, at a characteristic 
time scale of one or two weeks or so. Moreover, the 
diamagnetization loops in the newly segregated phase are very weak, presumably due to
the existence of additional inhomogeneity down to a nanometer scale.
Our results also show that the optimally doped {\rm Bi$_2$Sr$_2$CaCu$_2$O$_{8+x}$}
crystals seem capable to remain in a single phase with time. 
\end{abstract}

\maketitle

That the cuprates are of a single phase at a given 
temperature and a carrier concentration is 
an essential assumption in the widely-used phase diagrams, 
\cite{phasediagrams1,phasediagrams2}
based on which many experimental aspects in the 
normal and superconducting states 
have been correlated and discussed.
However, it is often a primary 
difficulty to obtain single crystals with high structural 
perfection and doping uniformity, yet with large enough size 
suitable for many experiments. 
Besides the electronic phase separation,\cite{stripe1,stripe2}
chemical phases separation at mesoscopic or even larger scales, 
often related to the inhomogeneous distribution or domain 
formation of oxygen vacancies or dopants, 
has also been observed in many families of the 
cuprates,\cite{review1,review2} for examples,
in {\rm La$_2$CuO$_{4+x}$}, \cite{LCO4x1,LCO4x2,LCO4x3,LCO4x4}
in lightly Sr-doped {\rm La$_{2-x}$Sr$_x$CuO$_4$},\cite{LSCOcho}
in {\rm Nd$_{2-x}$Ce$_x$CuO$_{4-y}$}, \cite{Skelton}
and in underdoped {\rm YBa$_2$Cu$_3$o$_{7-x}$} where   
vacancy ordering domains of different types
form at the chain layers.\cite{YBCO1,Chen_review}
Phase inhomogeneity has also been observed 
in {\rm Bi$_2$Sr$_2$CaCu$_2$O$_{8+x}$} by 
low temperature scanning tunneling microscopic studies.
\cite{Pan,Lang} With all these observations, however, 
one still cannot rule out 
the possibility that the phase separation is an extrinsic effect, 
avoidable if the crystals are grown up 
and annealed properly. In this paper, through 
investigating the time evolution of the diamagnetization signal, 
we show that the oxygen dopants in originally single-phased
overdoped {\rm Bi$_2$Sr$_2$CaCu$_2$O$_{8+x}$} single crystals tend to 
redistribute spontaneously at room temperature, which eventually 
and unavoidably leads to the occurrence of phase separation and 
inhomogeneity in the titled compound.

The {\rm Bi$_2$Sr$_2$CaCu$_2$O$_{8+x}$} single crystals used in 
this experiment were proven to be of high structural quality and 
stoichiometric purity with the resolution limits
of our X-ray diffraction and energy dispersion 
X-ray studies. To further guarantee the uniformity of the crystal, 
only those tiny pieces of crystalline sheets (typically 
0.5 mm $\times$ 0.5 mm $\times$ 0.05 mm, $\sim$ 10$^{-2}$ 
mm$^3$ in volume) cleaved and cut from the center of larger uniform 
ones were used for investigation. Each crystal was carefully 
annealed to maintain a certain and uniform level of overdoping. 
Then, its susceptibility was measured in a weak ac 
magnetic field (0.3 to 60 G) using a high-sensitivity vibration-sample 
magnetometer specially designed for measuring 
sub-millimeter-sized specimens.\cite{setup}
This magnetometer uses a double-synchronous 
detection technique, with a high-frequency ac magnetic field at 10 kHz 
(at which the {\rm Bi$_2$Sr$_2$CaCu$_2$O$_{8+x}$} crystal is fully 
penetrated in the normal state), and a low-frequency vibration of 
the sample at 4.4 Hz. It reaches a resolution of $1\times 
10^{-8}$ emu in susceptibility (or $3\times 10^{-10}$ 
emu in magnetic moments) for a sample of volume 1 mm$^ 3$ in a field of 30 G.
With such a sensitivity we can resolve any minority 
superconducting phase whose total volume is as small as 10$^{-7}$ mm$^3$, 
at the same time without suppressing the superconductivity. Our 
measurement on optimally doped {\rm Bi$_2$Sr$_2$CaCu$_2$O$_{8+x}$} 
crystals reveals a single sharp transition (will 
be shown in Fig. 5), which proves the reliability of this home-made
setup on the one hand and the high quality of the pristine crystals 
used in this experiment on the other hand.

Figure 1 shows the superconducting transitions of two overdoped 
{\rm Bi$_2$Sr$_2$CaCu$_2$O$_{8+x}$} single crystals, samples \#1 
and \#1', measured immediately after an annealing treatment in 
O$_2$ atmosphere at 550 $^o$C for 52 hours. Sample \#1 is measured in 
the configuration of B$\bot$c, whereas sample \#1' is a small 
piece cut from sample \#1, in order to be mounted into the 
quartz-tube sample holder in the B$\bot$ab configuration. 
The 10\% to 90\% transition 
width for sample \#1' is $\sim$ 1 K in a field of 0.6 G perpendicular 
to the ab-plane, which demonstrates that an overdoped crystal could
be in a uniform single phase. 

For the case of B$\bot$c, the 
diamagnetization signal at low temperatures reflects the nature 
of intrinsic Josephson couplings along the c-direction. 
\cite{cJosephson}  To distinguish the transition regime from this 
Josephson coupling regime, we find it helpful to plot the data in 
a logarithmic scale (the inset of Fig. 1). In such a plot the 
temperature dependence of the susceptibility is very steep in the 
transition regime but relatively ``flat" in the Josephson coupling 
regime. The crossover between the two regimes takes place at $\sim$ 
82.5 K for sample \#1.

Although the overdoped crystals show a single superconducting 
transition if measured immediately after an annealing treatment, 
multiple transitions at different levels are always seen after the crystals 
have been kept at room temperature for a period of time. 
In our experiment, nine overdoped crystals have been examined,
seven of them eventually 
exhibited a multiple transition pattern, and two of them underwent 
significant broadening in transition width, after they have been 
stored for a certain period of time. 

Figure 2 shows the time evolution of the transition pattern for sample \#2.
This crystal was annealed in O$_2$ atmosphere at 450 $^o$C for 48 hours. 
It was afterwards kept at room temperature in a dry air environment 
as protected by silica gel, which is a common way of sample storage 
in most of the laboratories. In a three-month period
the susceptibility of the crystal in the B$\bot$c configuration 
was measured for several times. From the panel (b) of Fig.2 it can be 
seen that the 88 K phase grew up significantly with time, and reached 
$\sim$ 10\% in diamagnetization at the end of the three-month 
period. 

The measurement performed in the B$\bot$ab configuration on sample \#2'
(which is a small piece taken from sample \#2 at the end
of the three-month period) also revealed the formation of a very pronounced 
secondary phase, as shown in the panel (a) of Fig. 2. 

The growing up in volume of a higher-T$_c$ phase should reflect
the change in local oxygen content and/or arrangement in the crystals.  
Only with such a mechanism can then the change be reversible, i.e.,
a single sharp transition can be 
restored by high-temperature annealing --- which 
was exactly the case for our crystals. Our results therefore reveal that 
the balanced oxygen content in overdoped 
{\rm Bi$_2$Sr$_2$CaCu$_2$O$_{8+x}$} is a function of the temperature
and tbe oxygen partial pressure of the crystal's environment. The 
nearly balanced oxygen content in a crystal, as maintained by high-temperature
annealing in O$_2$-atmosphere, must become over saturated as the crystal is
cooled down to room temperature. 
Therefore, the oxygen dopants tend to redistribute and even to escape
from the crystal, which finally creates the higher-T$_c$ phase.
For this reason, the newly formed phase is likely near the edge of 
the crystals. It should occupy a substantially large volume,
in order to cause an apparent diamagnetic signal of 
$\sim$10\% for sample \#2, and up to $\sim$ 50\% for sample \#2'. 

The distinct multiple-transition pattern indicates 
the formation of two types of domains in the ab-plane 
of the crystal, each having a different oxygen content.
The mechanism of forming such domains might be similar
to that of the vacancy ordering domains
in underdoped {\rm YBa$_2$Cu$_3$O$_{7-\delta}$}.\cite{YBCO1,Chen_review}
It has been shown through 
numerical simulations \cite{numerical} that an ordered distribution of 
oxygen vacancies commensurable with the lattice, i.e., forming some 
types of staging, is energetically more favored than a random 
distribution. 

A non-trivial mobility for the oxygen dopants at room temperature
is another key factor for the occurrence of spontaneous phase separation. 
It has been shown in {\rm La$_2$CuO$_{4+x}$} that the oxygen mobility 
is linked to the lattice imperfections, e.g., it is strongly 
enhanced by the presence of planar defects.\cite{mobility} For 
{\rm Bi$_2$Sr$_2$CaCu$_2$O$_{8+x}$} samples, 
the existence of incommensurate modulation along the 
b-axis \cite{incommensurate1,incommensurate2} 
might in some way influence the oxygen mobility, which needs to be further studied.

Figure 3 shows the field dependence of the 
transition pattern of sample \#2' three months after the annealing, 
and sample \#2 24 days after the annealing. Two transitions 
can be recognized in weak magnetic fields, one onsets at $\sim$ 
81.6 K and the other above 86 K. The nucleating and growing up of the 
superconducting domains below 86 K build up substantially large 
or enormous amount of diamagnetization loops in the ab-plane of 
the crystal. However, these loops are very weak, as they can be 
drastically suppressed by a field of $\sim$ 1 Gauss, and almost 
washed out by a field of a few tens Gauss. Such a sensitive field 
dependence should come from a Josephson-like weak superconductivity, 
reflecting the existence of additional inhomogeneity 
in the ab-plane of the newly segregated phase
down to the coherence length scale.
\cite{Other_field_dependence}

The diffusion process of the dopants to a lower-energy distribution
is presumably driven by the thermal energy which is significantly lower
at room temperature compared to that in high temperature annealing.
This factor makes the diffusion process hardly completed, givn rise to the
additional inhomogeneity in the ab-plane.
This picture seems consistent with the recent 
observation of energy gap inhomogeneity at nanometer scale using 
scanning tunneling microscopy.\cite{Pan,Lang} With the best single 
crystals they obtained, Pan and co-workers found that the 
superconducting energy gap of 
{\rm Bi$_2$Sr$_2$CaCu$_2$O$_{8+x}$} compound
is inhomogenous from place to place down 
to the nanometer scale, though the lattice is highly perfect. 

Finally, Fig. 5 shows the transition 
patterns of two optimally doped {\rm Bi$_2$Sr$_2$CaCu$_2$O$_{8+x}$} 
crystals (the 88 K phase) 15 days (sample \#3) and 
six months (sample \#4) after being annealed in Ar atmosphere
(600 $^o$C, two weeks). These crystals are from the same batch 
as sample \#1 and \#2. It seems that the tendency of spontaneous 
phase separation at room temperature is much insignificant for 
optimally doped crystals, as their transition pattern remains nearly 
the same after the six months period, and shows no significant field 
dependence at a level of a few Gauss. It indicates that 
the oxygen dopants in optimally doped crystals are near to
a minimum energy distribution which is homogenous 
down to the coherence length scale.

\begin{acknowledgments}
One of us (LL) thanks H. H. Wen and S. H. Pan for helpful 
discussion. This work was supported by the National Science 
Foundation of China, and by the National Center for R \& D of 
Superconductivity, China.
\end{acknowledgments}


\begin{figure} 
\caption{
The superconducting transition pattern of overdoped 
{\rm Bi$_2$Sr$_2$CaCu$_2$O$_{8+x}$} single crystals measured 
immediately after an annealing treatment in O$_2$ atmosphere 
at 550 $^o$C, showing a single and sharp transition. The inset is a 
semi-logarithmic plot of the data in the B$\bot$c configuration, 
which helps to distinguish the existence of any secondary superconducting 
phase in this configuration. Sample \#1' is a small 
piece cut from sample \#1.
}
\end{figure}  

\begin{figure} 
\caption{The time-evolution transition patterns of overdoped 
{\rm Bi$_2$Sr$_2$CaCu$_2$O$_{8+x}$} single crystals. A higher 
$T_c$ phase gradually grows up as the crystals are 
kept at room temperature in a dry-air atmosphere. At the end 
of the three-month period, the newly segregated phase caused an 
apparent diamagnetization volume of $\sim$ 10\% for the case of B$\bot$c 
(panel (b)), and up to $\sim$ 50\% for B$\bot$ab (panel (a)) at 82 K.}
\end{figure}  

\begin{figure}  
\caption{The field dependence of the multiple transitions 
of overdoped {\rm Bi$_2$Sr$_2$CaCu$_2$O$_{8+x}$} single crystals. 
Panels (a) and (b) are the susceptibility in the B$\bot$ab and 
B$\bot$c configurations, respectively. Panels (c) and (d) are the 
semi-logarithmic plots of the data in panels (a) and (b). Panels 
(e) and (f) displays the details of the onset-transition regions. } 
\end{figure}  

\begin{figure}  
\caption{The transition patterns of two optimally doped crystals 
of {\rm Bi$_2$Sr$_2$CaCu$_2$O$_8$} 15 days (sample \#3) and six 
months (sample \#4) after being annealed in Ar atmosphere at 600~$^O$C. 
Their transition patterns remain nearly the same, and shows no 
significant field dependence at a level of a few Gauss.}
\end{figure}  

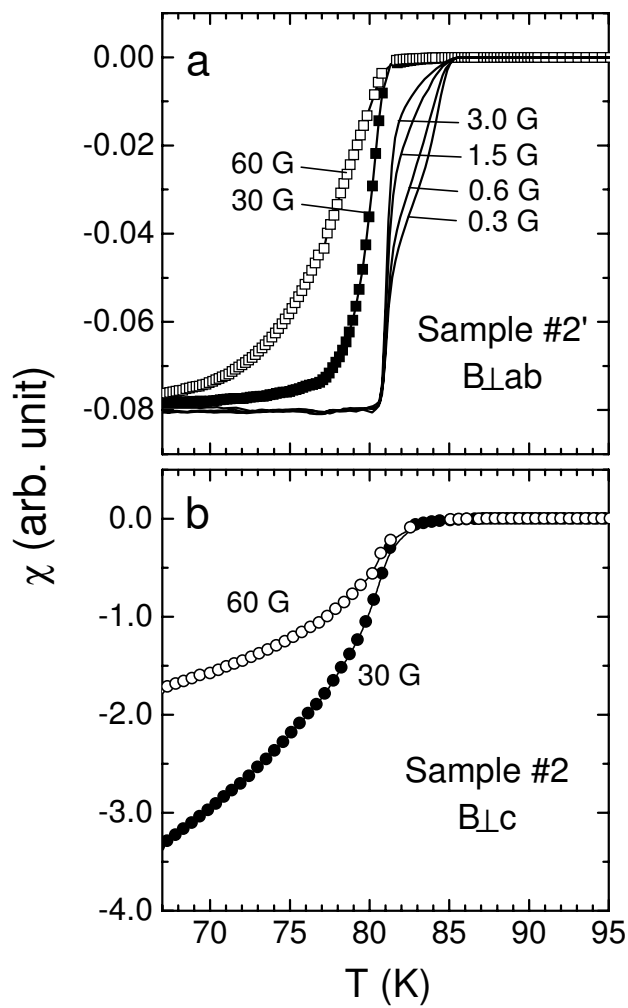
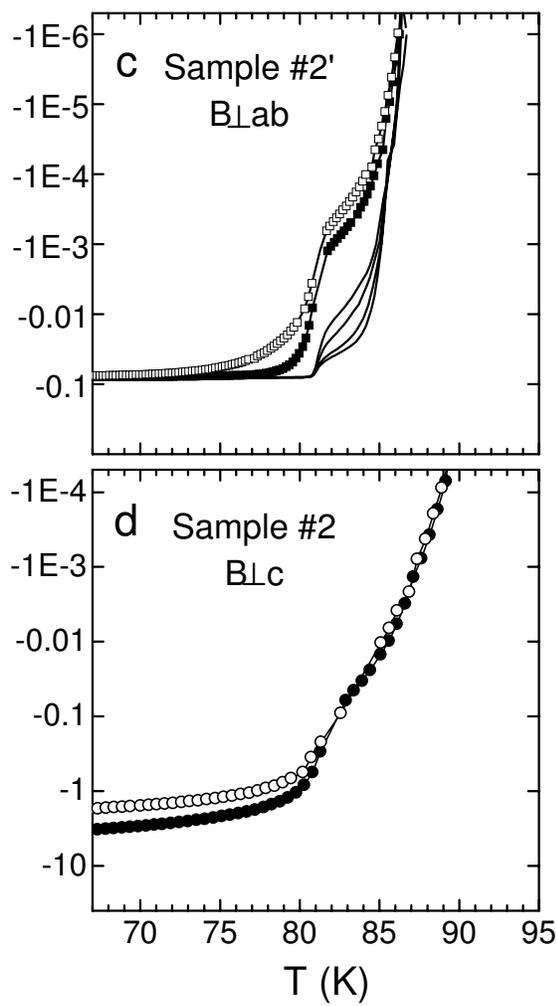
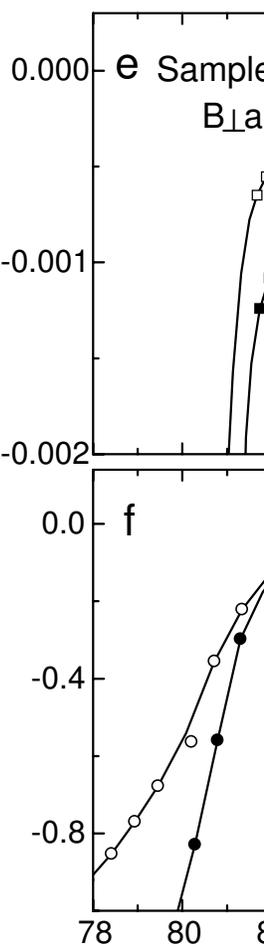
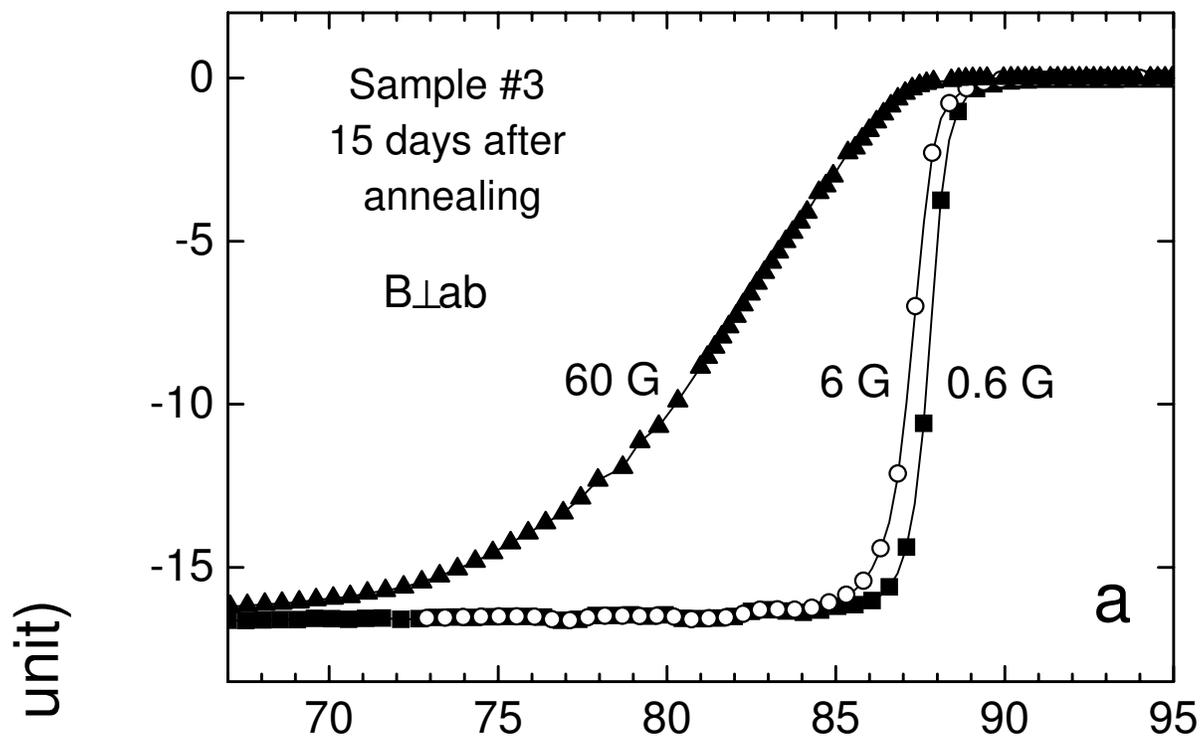
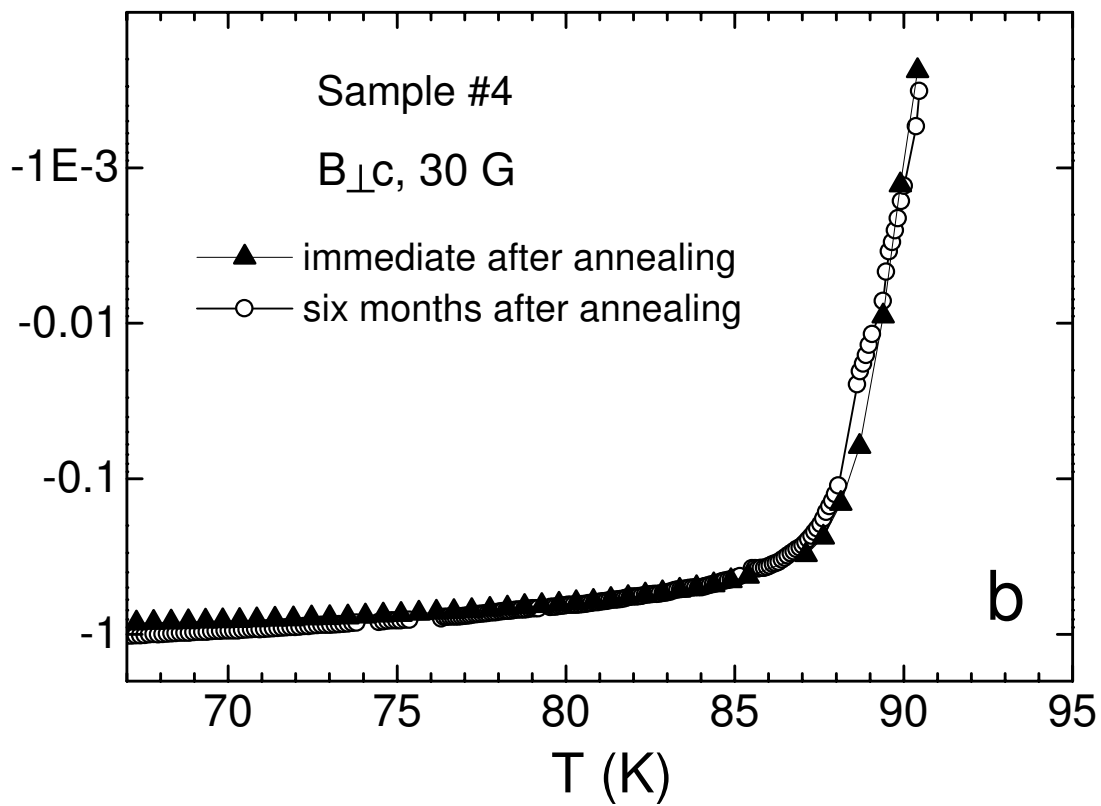
\end{document}